\documentclass[12pt,numbers]{elsarticle}
\makeatletter
\def\ps@pprintTitle{%
 \let\@oddhead\@empty
 \let\@evenhead\@empty
 \def\@oddfoot{}%
 \let\@evenfoot\@oddfoot}
\makeatother
\usepackage{latexsym}

\usepackage[latin2]{inputenc}

\begin{document}

\title{An Intuitive Graphical Webserver for Multiple-Choice Protein Sequence Search} 

\author[p,u]{Dániel Bánky}
\ead{banky@pitgroup.org}
\author[p]{Balázs Szalkai}
\ead{szalkai@pitgroup.org}
\author[p,u]{Vince Grolmusz\corref{cor1}}
\ead{grolmusz@pitgroup.org}
\cortext[cor1]{Corresponding author}

\address[p]{Protein Information Technology Group, Eötvös University\\ Pázmány Péter stny. 1/C, H-1117 Budapest, Hungary}
\address[u]{Uratim Ltd., H-1118 Budapest, Hungary}

\begin{abstract}
 Every day tens of thousands of sequence searches and sequence alignment queries are submitted to webservers. The capitalized word "BLAST" become a verb, describing the act of performing sequence search and alignment. However, if one needs to search for sequences that contain, for example, two hydrophobic and three polar residues at five given positions, the query formation on the most frequently used webservers will be difficult. Some servers support the formation of queries with regular expressions, but most of the users are unfamiliar with their syntax.
Here we present an intuitive, easily applicable webserver, the Protein Sequence Analysis server, that allows the formation of multiple choice queries by simply drawing the residues to their positions; if more than one residue are drawn to the same position, then they will be nicely stacked on the user interface, indicating the multiple choice at he given position. This computer-game like interface is natural and intuitive, and the coloring of the residues makes possible to form queries requiring not just certain amino acids in the given positions, but  also small nonpolar, negatively charged, hydrophobic, positively charged, or polar ones. 
The webserver is available at http://psa.pitgroup.org.
\end{abstract}

\maketitle

\section{Introduction}

There are numerous webservers available that make possible searching protein databases for regular expressions, e.g.,:

\noindent Scansite http://scansite.mit.edu/dbsequence\_reg.html \cite{Obenauer2003}, Sequence motif search at RCSB PDB \cite{PDB-base}, or the Sequence Searcher \cite{Marass2009}. However, all of these servers require some knowledge of the syntactics of regular expressions, and that syntax are still not widely known for molecular biologists and medical researchers worldwide. 

In order to present an easy-to-understand, self-standing, and highly intuitive user interface, we developed a graphic webserver and the underlying program, the Protein Sequence Analyzer, that is available at http://psa.pitgroup.org.

\section{Results and Discussion:}
The webserver makes possible creating a graphical query-sequence that may contain any subset of the amino acids at any positions of the query, including wild cards "?" and "*", meaning "any amino acid" or "any subsequence of amino acids", respectively. The input of the query sequence is done by grabbing one of the amino acid codes or wild cards by your mouse or on your touch screen, and move it to the block with the yellow cross. 

If more than one residue is allowed on the same slot, the additional amino acid codes can be added by dragging it to the existing slot: as a result, the slot becomes larger, and the new residue code will be stacked under the previous ones.

If a wildcard is dragged to a collection of residue codes in a slot, the wildcard will be substituted for all the residue codes. 

The slots can be re-ordered by grabbing them at their purple handle on the top. A residue code can be deleted simply by clicking on it. If all the amino acid codes are deleted from a slot, then a slot itself will disappear.

The amino acid residues are colored by their physico-chemical properties in order to help the users to form queries that contain amino acids with certain properties. Namely, the color codes have the following correspondence to amino acid residue properties \cite{Lesk}:
\begin{itemize}
\item Orange: small non-polar
\item Red: negatively charged
\item Green: hydrophobic
\item Blue: positively charged
\item Pink: polar.
\end{itemize}

For specifying the search domain, the user may choose the Swiss-Prot and the TrEMBL subsets of the UniProt database, or even the whole UniProt \cite{Consortium2010}.

The processing time of the typical query is around half a minute on Swiss-Prot, but it strongly depends on the complexity of the query and the server load. 

The results are returned in the browser, and also as a downloadable Fasta file. 

The result sheet contains the query translated to a regular expression (for archiving the query), and the first 100 hits. The hits are identified by their UniProt accession numbers and the subsequence found is highlighted by red letters.

For advanced users a direct input of the regular expressions is also possible by choosing the "Advanced version" option.

{\bf Translation to regular expression:} The PSA server can also be used for translating the computer-game-like query to a regular expression: If the query is formed, then, by hitting "Advanced version" the regular expression, corresponding to the query, is returned.
\begin{figure*}
\centering
\includegraphics[width=165mm]{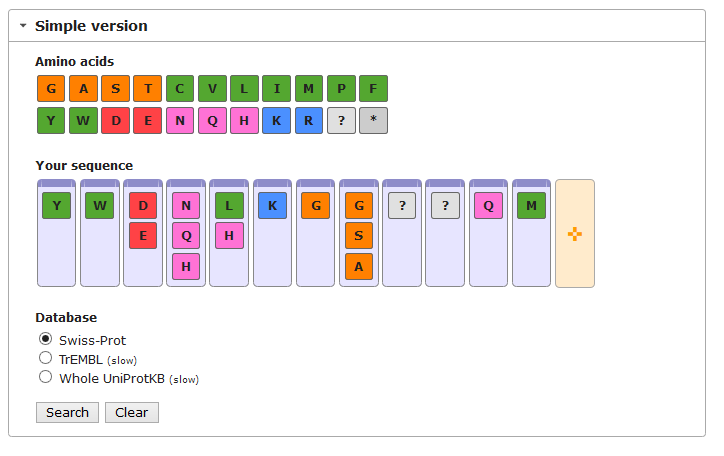}
\caption{The intuitive user interface of the PSA server. The residues need to be dragged from the top block to either to the yellow cross at the right end of the query block, or to just any slots, already containing one or more residue codes. Color codes describe the physico-chemical properties of the amino acids. Adding a wildcard ("?" or "*") to an existing slot will delete the codes from the slot and substitute the wildcard. The result sheet contains the regular expression generated from the query, and the result sequences, with their UniProt accession numbers in two formats: in fasta for downloading and in the browser, with the hits highlighted with color. }
\end{figure*}

\section{Conclusions:} We presented the PSA webserver with an easy-to-manage, intuitive user interface that makes possible to form multiple-choice queries with dragging symbols to slots in a sequence. The webserver works with mice on desktop and laptop computers and touch-screens on mobile devices. The result is downloadable in fasta format, and it is also returned in a highlighted html version for quick review.

\end{document}